\documentclass{emulateapj}
\pdfoutput=1

\usepackage{amsmath}
\usepackage{amssymb}
\usepackage{color}
\usepackage{dcolumn}
\usepackage{graphicx}
\usepackage[utf8]{inputenc}
\usepackage{latexsym}
\usepackage{times}

\newcommand{\scri}{\ensuremath{\mathcal{J}^+}}

\newcommand{\Ylm}[2]{{}_{-2}Y_{#1 #2}}
\newcommand{\news}{\ensuremath{\mathcal{N}}}

\definecolor{rgb_red}{rgb}  {0.7, 0.0, 0.0}
\definecolor{rgb_green}{rgb}{0.0, 0.6, 0.0}
\definecolor{rgb_blue}{rgb} {0.0, 0.0, 0.6}

\begin{document}

\title{Gravitational memory in binary black hole mergers}

\author{Denis Pollney}
\affil{Departament de F\'isica,
  Universitat de les Illes Balears,
  Palma de Mallorca, E-07122 Spain
}
\email{denis.pollney@uib.es}

\author{Christian Reisswig}
\affil{
  Theoretical Astrophysics Including Relativity, 
  California Institute of Technology,
  Pasadena, CA 91125, USA
}
\email{reisswig@tapir.caltech.edu}

\begin{abstract}
  In addition to the dominant oscillatory gravitational wave signals
  produced during binary inspirals, a non-oscillatory component arises
  from the nonlinear ``memory'' effect, sourced by the emitted
  gravitational radiation. The memory grows significantly during the
  late inspiral and merger, modifying the signal by an almost
  step-function profile, and making it difficult to model by
  approximate methods. We use numerical evolutions of binary black
  holes to evaluate the nonlinear memory during late-inspiral, merger
  and ringdown. We identify two main components of the signal: the
  monotonically growing portion corresponding to the memory, and an
  oscillatory part which sets in roughly at the time of merger and is
  due to the black hole ringdown. Counter-intuitively, the ringdown
  is most prominent for models with the lowest total spin.
  Thus, the case of maximally spinning black holes \emph{anti-aligned} to the
  orbital angular momentum exhibits the highest
  signal-to-noise (SNR) for interferometric detectors. The largest memory
  offset, however, occurs for highly spinning black holes, with an estimated
  value of $h^{\rm tot}_{20}\simeq 0.24$ in the maximally spinning
  case. These results are central to determining the detectability of
  nonlinear memory through gravitational wave interferometers and 
  pulsar timing array measurements.
\end{abstract}


\maketitle

\section{Introduction}

Accelerating massive bodies generate gravitational waves (GWs), a
potentially rich source of astrophysical information which a number of
large experiments have been designed to mine over the next decade.
The principle sources for the interferometric detectors (LIGO, Virgo,
LISA) are orbiting bodies, for which the dominant signal is
oscillatory. However, it has long been known that GWs will also contain
non-oscillatory features, resulting from a net change in time
derivatives of the multiple moments of the
system~\citep{1974SvA....18...17Z, 1987Natur.327..123B}.  A nonlinear
contribution also results from the interaction of the waves with
themselves~\citep{Payne:1983, Christodoulou:1991cr,
  Blanchet:1992br}. Evaluating these modes involves an integral in
time, requiring knowledge of the entire past history of the
spacetime. As such, they have been given the name ``memory'' or
``hereditary'' components of the GW.

The nonlinear memory is sourced by the emitted GWs.
For a relativistic binary inspiral, the typical profile is of a
slow growth over time which sees a rapid increase during the late
inspiral and merger, and reaches a constant value as the merger
remnant rings down and ceases to emit. The non-oscillatory nature of
the memory suggests that it will not be a dominant feature in
interferometric detectors, for which drifts in the background metric
are factored out. However, the step induced during merger may be
observable in pulsar timing measurements, as has
been noted in a recent set of papers~\citep{Pshirkov:2009ak,
  vanHaasteren:2009fy, Seto:2009nv}.

Post-Newtonian (PN) calculations provide an accurate estimate of the
memory to within a few orbits of the merger~\citep{Kennefick:1994nw},
however fail to model the important merger phase where the effect is
largest. In a series of papers, ~\cite{Favata:2008ti, Favata:2008yd,
  Favata:2009ii, Favata:2010zu} has included an estimate of the merger
contribution for binary black holes (BHs) by using an
effective-one-body (EOB) model which was tuned using the results of
numerical simulations. Recent progress in numerical models of BH
spacetimes provides the opportunity to measure the effect directly,
though as outlined in~\cite{Favata:2008yd}, due to the low amplitude
of the memory, and systematic error in standard techniques of wave
measurement in numerical relativity, the problem is a challenging one.

In this paper, we apply a newly developed evolution
code~\citep{Pollney:2009yz, Pollney:2009ut} and wave extraction
techniques~\citep{Reisswig:2009us, Reisswig:2009rx} to carry out fully
relativistic binary BH simulations to directly measure the nonlinear
GW memory through inspiral, merger and ringdown in a gauge-invariant
and mathematically unambiguous notion.  We concern ourselves with
equal-mass, spinning binaries. For these models, the memory is
contained in the $\Ylm{\ell}{0}$ spin-weighted spherical harmonic
components of the GW strain, $h^+$. These modes, previously studied in
the context of head-on collisions~\citep{Anninos94b}, exhibit two
dominant effects. The first is a non-oscillatory term which rises as
the inspiral progresses, and is associated with the memory. During the
merger, an oscillatory signal is superposed onto these modes, induced
by the ringdown of the BH remnant.  After the ringdown has subsided,
the modes are offset from their original value. We find that for the
dominant $(\ell,m)=(2,0)$ memory mode, this offset is largest in the
case of the merger of aligned, maximally spinning BHs, as might be
expected given that these are the strongest gravitational emitters
\citep{Reisswig:2009vc}.  Interestingly, we observe the strongest
oscillatory ringdown signal in the case of lowest total spin
(maximally spinning BHs anti-aligned with the orbital angular
momentum). As such, and as opposed to the observability of the
dominant $(\ell,m)=(2,2)$ mode, these models produce a more visible
$(\ell,m)=(2,0)$ mode to interferometric detectors. We conclude by
estimating the strain offset which would be visible to a pulsar timing
array.

\section{Numerical methods}\label{sec:code}

We integrate the BSSNOK formulation of the vacuum Einstein equations
(see, e.g.,~\citealt{Alcubierre:2008}) numerically, using the
\texttt{Llama} evolution code~\citep{Pollney:2009ut, Pollney:2009yz}
within the Cactus framework~\citep{Goodale02a}. The evolution
equations are discretized via
finite differences in space on a grid composed of
multiple patches with locally adapted coordinates in the wave
zone. Adaptive mesh-refinement is implemented via the Carpet grid
driver~\citep{Schnetter-etal-03b} on the central patch
surrounding the BHs.  Importantly for the measurement of small
non-oscillatory features such as the memory, the artificial grid outer
boundary is causally disconnected from the wave measurements.

We determine initial data for binary BHs by the conformal puncture
data method~\citep{Brandt97b}, and evolve them using the ``moving
puncture'' approach~\citep{Baker:2005vv, Campanelli:2005dd} whereby
the choice of gauge \citep{Alcubierre02a} prevents the spacetime
slicing from encountering the curvature
singularity~\citep{Hannam:2006vv}.

Crucial for determining the memory, we compute the GW signal of the
source using the method of Cauchy-characteristic extraction (CCE),
which computes unambiguous and \textit{coordinate invariant}
signals at future null infinity,
\scri, corresponding to a detector far removed from the
source~\citep{Bishop98b, Babiuc:2005pg, Winicour05,
 Reisswig:2009us, Reisswig:2009rx}.  By this technique, metric data
is collected on a world-tube at finite radius, and a formulation of
the full Einstein equations along null hypersurfaces is used to
transport the signal to~\scri.

We measure two quantities that encode the gravitational signal at
\scri. The first is the Newman-Penrose scalar $\psi_4$, the Weyl
curvature component with slowest falloff in asymptotically flat
spacetimes~\citep{Newman62a}. 
Alternatively, we
measure the Bondi ``news''~\citep{Bondi62, Sachs62}, defined by
\begin{equation}
  \news = -\Delta \bar\sigma,
\end{equation}
where $\Delta = l^a\nabla_a$ is a derivative operator defined on an
outgoing null geodesic, $l^a$, and $\sigma$ is the Newman-Penrose
shear scalar~\citep{Newman62a}. These scalars, $\psi_4$ and $\news$,
can be evaluated directly from the local curvature on the sphere at
\scri and in the naturally defined gauge of an asymptotic observer~\citep{Bondi62, Sachs62}. 
To calculate the GW strain $h$ which is observed by a detector, these
quantities need to be numerically integrated in time:
\begin{equation} \label{eq:h_from_N}
  h = h_+ - ih_\times 
    = \int^t_{-\infty} dt' \news\,
    = \int^{t}_{-\infty} dt^\prime 
      \int^{t^\prime}_{-\infty} dt^{\prime\prime} \psi_4\,.
\end{equation}

We expand $h$ in terms of a basis of spin-2 spherical harmonics
$\Ylm{\ell}{m}$. The dominant non-oscillatory hereditary
component is contained in the $(\ell,m)=(2,0)$ mode,
\begin{equation}
  r\,h_{20} \equiv \int_\Omega d\Omega ({}_{-2}\bar{Y}_{20}) h,
\end{equation}
where $\Omega$ is the sphere at~\scri and $r$ is the distance
to the source.

Since the simulations are necessarily of finite length,
the time integration of $\news$, Eq.~\eqref{eq:h_from_N}, leaves a
constant to be determined, corresponding to the initial value of the
signal at the start time of the simulation. For purely oscillatory
signals (e.g., $h_{22}$) we can fit this by adjusting the post
ringdown value to zero, or by some averaging procedure over multiple
inspiral wavelengths. This is not possible in the case of the
monotonically growing memory. Instead, we approximate an initial value
of the $h_{20}$ mode by matching to a PN calculation,
using the 3PN expansion for $h^{\rm mem}$ derived
in~\cite{Favata:2008yd}. 
For spinning binaries we incorporate 2.5PN spin
contributions to the frequency evolution~\citep{Blanchet:2006gy}.  The
initial offset of the numerical waveforms is performed by shifting its
amplitude so that it fits against the PN models, over an interval from
$t=-300M$ to $t=-200M$. For instance, the PN solution for $h_{20}$ in
the non-spinning case is plotted as a dotted line in the lower panel
of Figure~\ref{fig:memory-vs-time}. In Table~\ref{tab:memory-vs-spin},
we report both the overall amplitude offset of the ringdown signal in the
$h_{20}$ mode matched to the PN inspiral, as well as the offset from
the point $t=-300M$ which is computed entirely from numerical data.

\section{Characteristics of the memory modes}

\begin{figure}
  \includegraphics[width=1.\linewidth]{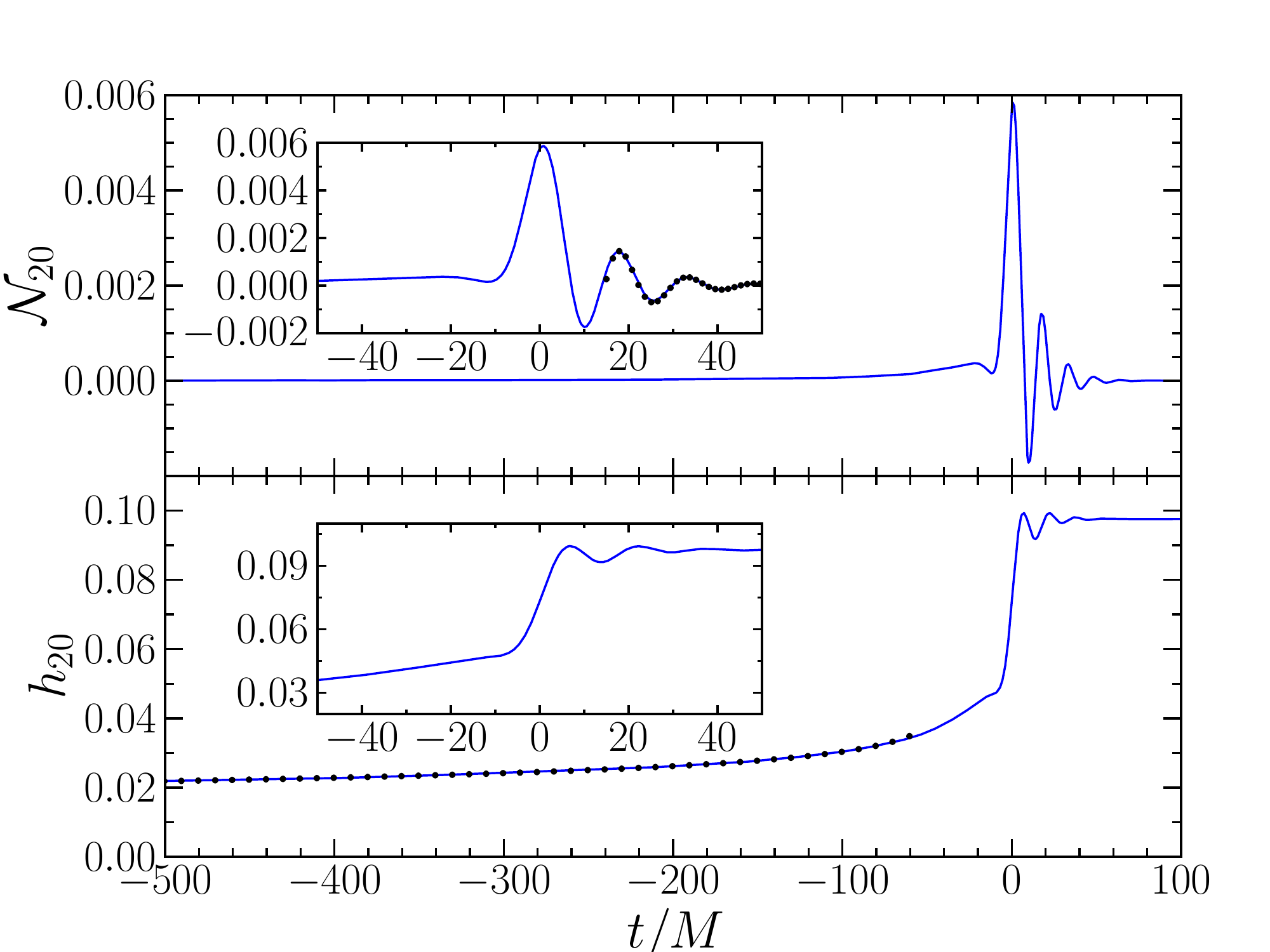}
  \caption{\textit{Upper panel:} the $(\ell,m)=(2,0)$ mode of the
    news, $\news$. \textit{Lower panel}: the strain, $h$ (integrated from $\news$). 
    The dotted line in the
    inset of the upper panel is the result of a fit to the ringdown
    QNM, Eq.~\ref{eq:qnm_fit}, while in the lower plot it
    denotes the 3PN memory derived in~\cite{Favata:2008yd}, used to
    determine the integration constant.
}
  \label{fig:memory-vs-time}
\end{figure}

In Figure~\ref{fig:memory-vs-time}, we show the $(\ell,m)=(2,0)$ mode
during late inspiral and merger for the case of an equal-mass,
non-spinning binary using data from the model
presented in~\cite{Pollney:2009yz} and evaluated at 
\scri~in~\cite{Reisswig:2009us}.  The upper panel shows the development of
the $(\ell,m)=(2,0)$ component of the news, $\news_{20}$, whereas the
lower shows its integral, $h_{20}$.  The time coordinate places $t=0M$
at the position of the amplitude peak of the dominant $(2,2)$ mode
(not shown), corresponding roughly to the time of merger. Two main
features are present. The first is a slow monotonic growth of the
$\news_{20}$ mode during the inspiral and plunge phase. This
corresponds to the non-oscillatory memory contribution. At $t\simeq
-20M$, there is a local maximum in $\news_{20}$, and shortly
afterward a prominent oscillatory signal sets in. This signal
represents the binary merger and subsequent BH ringdown. It is not
strictly an aspect of the GW ``memory'', as it is not dependent on the
entire past history of the spacetime, however it is superposed on the
growing memory signal.

The lower plot shows the strain, determined via
Eq.~\ref{eq:h_from_N}. The superposition of the memory and ringdown
signals results in a notable kink in the waveform near $t=0M$, and
exponentially decaying oscillation before reaching the steady-state
amplitude of
\begin{equation}
  h^\text{tot}_{20}=0.097\pm2\times10^{-3}.
\end{equation}
We measure approximately 4-th order convergence in
$\news_{20}$ when comparing three different grid resolutions for the
same model, and we estimate an error on the order of $1\times10^{-4}$
in $\news_{20}$. The much larger error of $2\times10^{-3}$ in
$h^\text{tot}_{20}$ is an upper limit of the estimate on the total
error composed of integration error (\ref{eq:h_from_N}) and error in the PN fit for all
of the binary models considered in this paper.

Perturbative results provide a prediction for the observed
quasi-normal modes (QNM) of $\news_{20}$~\citep{Berti:2009kk}.  We
perform a least-squares fit of the numerical results to a function of
the form
\begin{equation}
  f(t) = A \exp(-t / \tau) \cos(\omega_{20}  t + \phi),
  \label{eq:qnm_fit}
\end{equation}
over the ring-down portion, $t\in[30,80]$, with fitting parameters,
$A$, and $w_{20}$, $\tau$, $\phi$, and arrive at a frequency
\begin{equation}
  M_\text{f}\omega_{20} =  0.3940 \pm 2\times 10^{-4}\,,
\end{equation}
using the measured mass of the remnant of $M_{\rm f}=0.951764 \pm
20\times 10^{-6}$~\citep{Pollney:2009yz}.  \cite{Berti:2009kk} provide
a tabulated value of $M\omega_{20}^{\rm lit.}=0.393245$ for the $(2,0)$
mode of a BH with the measured dimensionless spin
$a_\text{f}=0.686923\pm 1\times 10^{-5}$. Thus, we find a difference
of
\begin{equation}
  |M_\text{f}\omega_{20}-M\omega_{20}^{\rm lit.}|\approx8\times 10^{-4}\,,
\end{equation}
representing an error of less than $0.2\%$.  We list the final
spin and the associated QNM frequencies of the binary systems for all
simulations performed in Table~\ref{tab:memory-vs-spin}.

Finally we have also examined $m=0$ modes for $\ell=4$
and $\ell=6$.  For this
model, the overall amplitudes of the final memory are $h_{40}^{\rm
  tot}=1.7\times10^{-3}$, and $h_{60}^{\rm tot}\approx
5.0\times10^{-4}$.  However, these modes 
are difficult to distinguish from the
numerical error, and we conclude that the higher order modes
contribute less than $2\%$ to the total memory.

\section{Memory from spinning binaries}
\label{sec:spin}

We have performed simulations using the code infrastructure
described in Section~\ref{sec:code} in the two-dimensional parameter space
of non-precessing equal-mass binaries with spin
that has also been studied in~\cite{Reisswig:2009vc, Rezzolla:2007xa}. 
We have focused on initial data configurations
along the two main axes in the space of individual dimensionless BH
spins $(a_1, a_2)$ aligned with the orbital angular momentum.
The first set of models have equal but
opposite spins, $a_1=-a_2$; The second have identical spins,
$a_1=a_2$.
The model details are listed in Table~\ref{tab:memory-vs-spin}.

\begin{figure}
  \includegraphics[width=1.\linewidth]{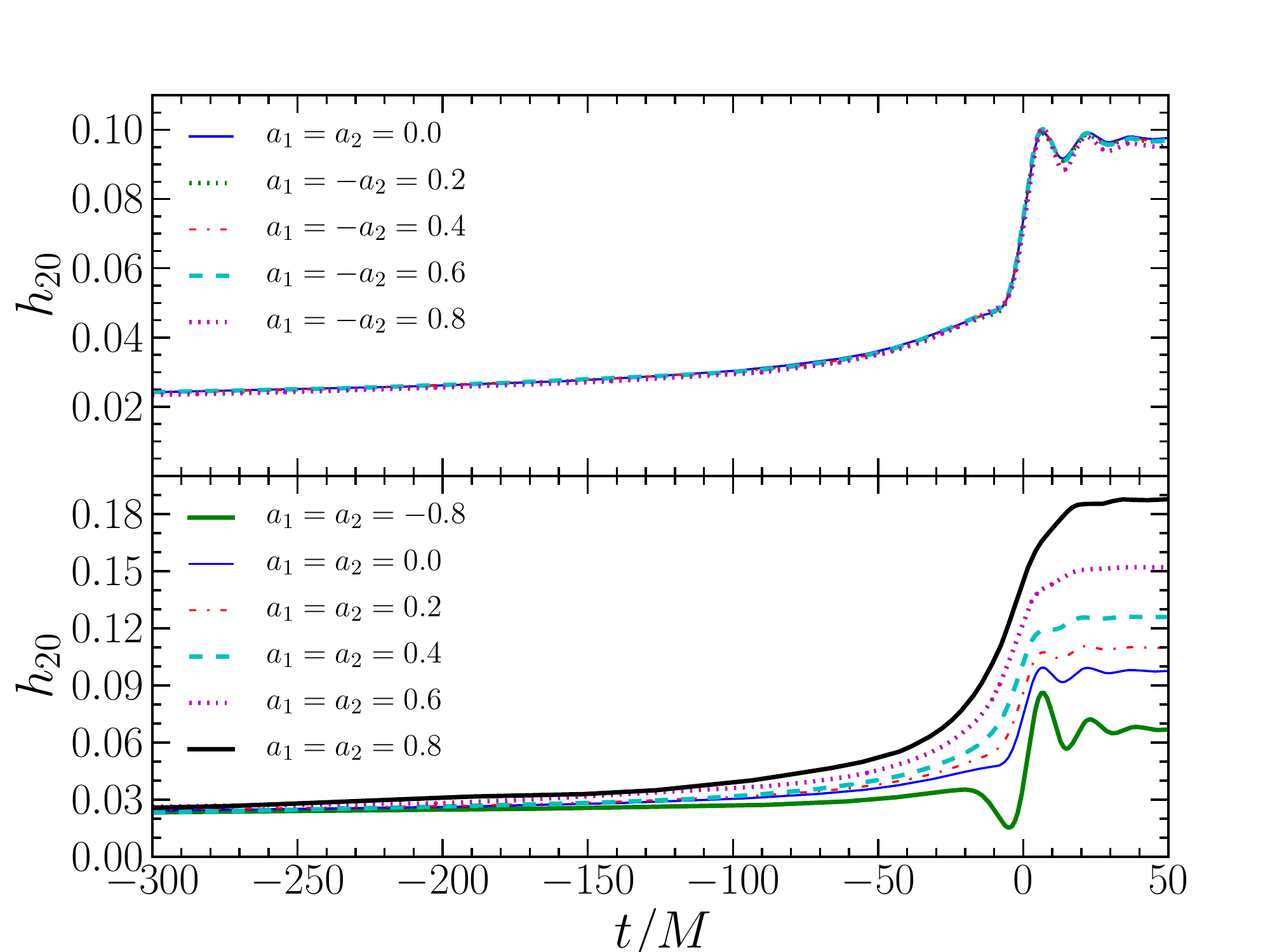}
  \caption{\textit{Upper panel:} the $(\ell,m)=(2,0)$ modes are
    similar for models with anti-aligned
    spins, $a_1=-a_2$.n
    \textit{Lower panel:} the memory offset grows with the total spin, 
    however the ringdown
    is most prominent for the anti-aligned spins, $(a_1,a_2)<(0,0)$.
  }
  \label{fig:memory-vs-spin}
\end{figure}

Figure~\ref{fig:memory-vs-spin} plots the evolution of the $h_{20}$
modes for the two sequences. The upper panel shows models for which
the spins of the individual BHs are anti-aligned, $a_1=-a_2$, so that
the spacetimes have the same total angular momentum. The nearly
identical lines are consistent with previous observations that the
$(\ell,m)=(2,2)$ modes also do not vary appreciably along this
direction of parameter space~\citep{Vaishnav:2007nm, Reisswig:2009vc},
as well as the radiated energies~\citep{Reisswig:2009vc}, at least not
to within the error-bars as stated in the latter reference.

The lower plot shows the $h_{20}$ modes for configurations with equal
and aligned spins, $a_1=a_2$.  The amplitude of the memory offset
increases with higher total spin. This is expected as the memory is
sourced by the emitted GWs which are known to be more energetic in the
higher spin models~\citep{Reisswig:2009vc, Lousto:2009mf}. The total
offsets, $h^\text{tot}_{20}$, are listed in
Table~\ref{tab:memory-vs-spin}.

We find that the amplitude of the ringdown is
largest for models with lowest total spin, $a_1=a_2=-0.8$. In these
cases, spin-orbit coupling increases the final angle of impact so that
the merger is more nearly head-on. The ringdown is suppressed in
the high spin $a_1=a_2=+0.8$ model. Fits to the QNM
frequencies are listed in Table~\ref{tab:memory-vs-spin} and show good
agreement with perturbative results. However, for simulations with
$a_1=a_2>0$ the weakness of the ringdown contributes to larger
variation in the estimates, particularly if the fitting interval is
varied, though still within $5\%$ of the analytic values.

\begin{figure}
  \includegraphics[width=\linewidth]{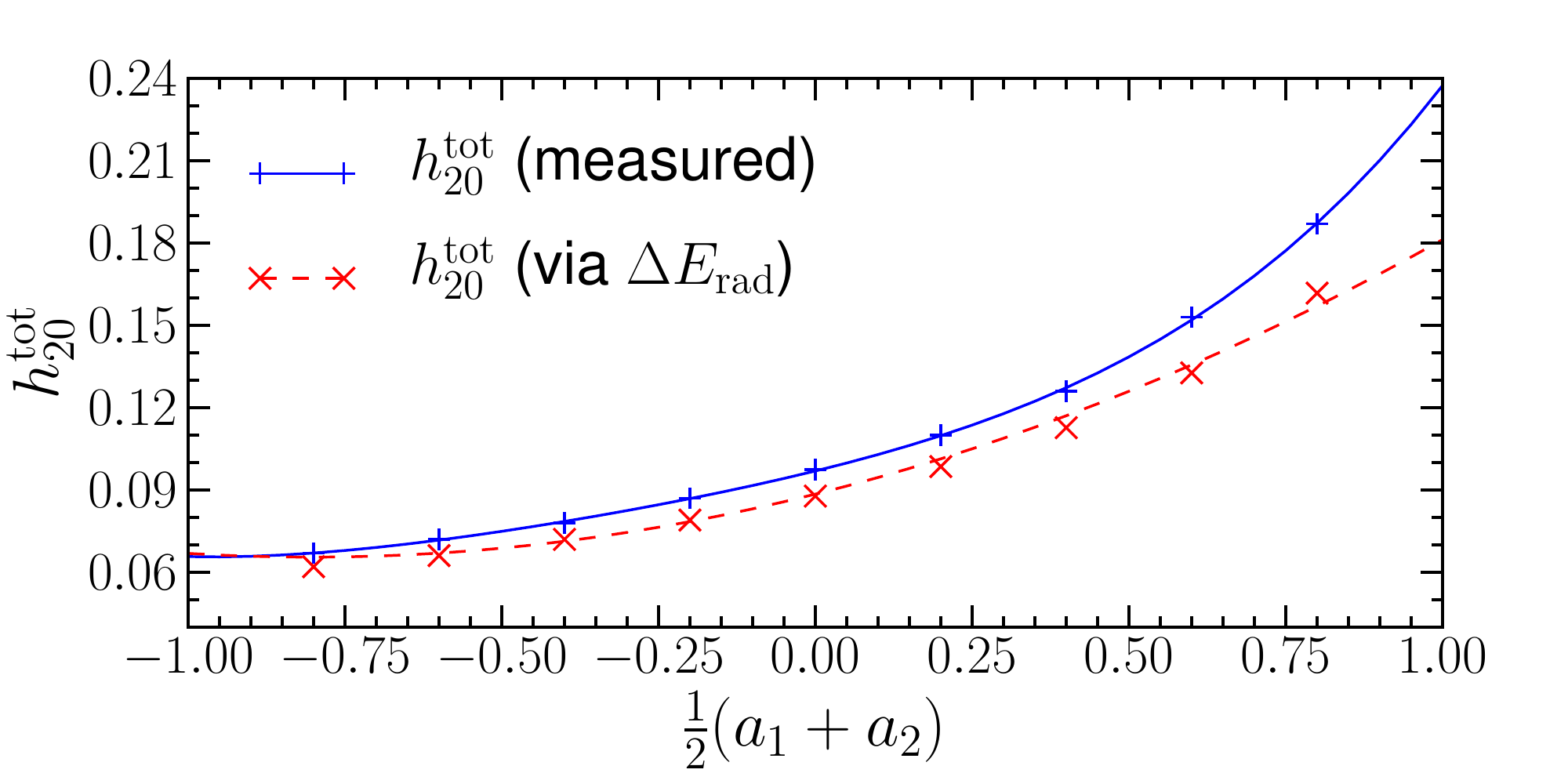}
  \caption{The total memory, $h^\text{tot}_{20}$, given by the offset
    of the $h_{20}$ after ringdown. The solid line is a quartic fit
    through the measured data ($+$) with integration constants
    determined by PN. This is compared with the estimate
    arising from the total radiated energy, via
    Eq.~\ref{eq:hmem_estimate} ($\times$, dashed line).
}
  \label{fig:h_fit}
\end{figure}

The memory is well estimated by a quartic polynomial in the total 
spin $a=(a_1+a_2)/2$,
\begin{align}
  h^\text{tot}_{20}(a) & r= 0.0969 + 0.0562\,a + 0.0340\,a^2 + 0.0296\,a^3
  \nonumber \\
  & + 0.0206\,a^4\,,
\end{align}
determined by a least-squares fit to the measured data.
Figure~\ref{fig:h_fit} plots $h^{\rm tot}_{20}$ as a function of spin
for the aligned models, $a_1=a_2$, as well as the leading order
multipole moment estimate resulting from the radiated energy
(see \cite{Favata:2008ti}, Eq.~(5))
\begin{equation}\label{eq:hmem_estimate}
  h^\text{mem}_+ \simeq \frac{\eta M h^\text{mem}}{384\pi R}
    \sin^2\theta (17 + cos^2\theta),
\end{equation}
with
\begin{equation}
  h^\text{mem} \simeq \frac{16\pi}{\eta}\left(
    \frac{\Delta E_\text{rad}}{M}\right).
\end{equation}
The measured radiated energy, $\Delta E_\text{rad}$, can also be
estimated using the quadratic fit developed in~\cite{Reisswig:2009vc},
verified here with the plotted data points corresponding to the new
and more accurate simulations (see Table~\ref{tab:memory-vs-spin}).
The radiated energy provides a reasonably good estimate of the memory
offset, though with a slight under-prediction.  The measured memory
values also appear to have a somewhat stronger dependence on the total
spin.  Extrapolating to the extremal $a_1=a_2=1$ case, we estimate a
maximum $h^\text{tot}_{20}\simeq 0.24$ in the case of aligned spins.
Fits to $\Delta E_\text{rad}$ for generic spin
configurations~\citep{Lousto:2009mf} suggest that this will indeed be
the maximum value for arbitrary spins.  Additional simulation in the
high-spin regime would be required to establish the accuracy of the
extrapolation.

\begin{table*}
\caption{
Parameters and measurements of physical quantities for the equal-mass
aligned-spin BBH models.
  From left to right, we list: initial dimensionless spins $a_1$ and $a_2$,
  the total memory $h_{20}^{\rm tot}$ accumulated from $T > -300M$ prior to merger as computed from NR simulations, and
  that accumulated from $T > -\infty$ by matching PN to numerical data, 
  the maximum SNR $\rho$ of the $(2,0)$ component attained
  over a range of masses ($M_{\rm opt} = 290 M_\odot$ for adLIGO, $M_{\rm opt} = 5.35\times10^6$
  for LISA) at luminosity distances $d=300
  {\rm Mpc}$ for adLIGO and $d=15.8 {\rm Gpc}$ ($z=2$) for LISA, where the
  bracketed numbers report the sky-averaged SNR,
  the total radiated mass $E_{\rm rad}$, including contributions
  from PN, as well as the final dimensionless spin $a_{\rm fin}$ of the
  remnant, 
  and finally, the measured QNM
  frequencies $M\omega_{20}$ and the error to perturbative calculations
  of~\cite{Berti:2009kk}.
  }
\vspace{-3mm}
\begin{center}
\begin{tabular}{|rr|cc|rr|r|r|cr|}
\hline
{$a_1$} 				&
{$a_2$} 				&
{$h_{20}^{\rm tot}|_{T\in[-300,100]}$}  &
{$h_{20}^{\rm tot}$}                    &
{$\rho^{({\rm mem})}_{\rm adLIGO}$} &
{$\rho^{({\rm mem})}_{\rm LISA}$}   &
{$E_{\rm rad}$~($\%$)}                  &
{$a_{\rm fin}$}                         &
{$M\omega_{20}$}                        &
{$|M\omega_{20}^{\rm lit.}-M\omega_{20}|$}  \\
\hline
\hline
$-0.2$ & $0.2$ & $0.073$ & $0.097$ & $5$ $(4)$ & $62$ $(51)$ & $4.84$ & $0.6866$ & $0.3932$ & $5\times10^{-4}$ \\    
$-0.4$ & $0.4$ & $0.073$ & $0.097$ & $5$ $(4)$ & $63$ $(52)$ & $4.85$ & $0.6865$ & $0.3932$ & $5\times10^{-4}$ \\
$-0.6$ & $0.6$ & $0.073$ & $0.097$ & $5$ $(4)$ & $65$ $(54)$ & $4.91$ & $0.6844$ & $0.3931$ & $3\times10^{-4}$ \\
$-0.8$ & $0.8$ & $0.072$ & $0.095$ & $5$ $(4)$ & $78$ $(64)$ & $4.94$ & $0.6832$ & $0.3978$ & $48\times10^{-4}$ \\
\hline
$-0.8$ & $-0.8$ & $0.044$ & $0.067$ & $11$ $(9)$ & $147$ $(121)$ & $3.42$ & $0.4247$ & $0.3824$ & $19\times10^{-4}$ \\
$-0.6$ & $-0.6$ & $0.049$ & $0.072$ & $9$ $(7)$ & $128$ $(106)$ & $3.66$ & $0.4919$ & $0.3824$ &  $6\times10^{-4}$ \\
$-0.4$ & $-0.4$ & $0.056$ & $0.078$ & $7$ $(6)$ & $103$ $(85)$ & $3.97$ & $0.5604$ & $0.3850$ & $11\times10^{-4}$ \\
$-0.2$ & $-0.2$ & $0.064$ & $0.084$ & $6$ $(5)$ & $81$ $(67)$ & $4.35$ & $0.6246$ & $0.3905$ & $11\times10^{-4}$ \\
$0.0$  & $0.0$  & $0.074$ & $0.097$ & $5$ $(4)$ & $63$ $(52)$ & $4.83$ & $0.6869$ & $0.3940$ & $8\times10^{-4}$ \\
$0.2$  & $0.2$  & $0.085$ & $0.110$ & $4$ $(3)$  & $47$ $(39)$  & $5.44$ & $0.7469$ & $0.3987$ & $12\times10^{-4}$ \\    
$0.4$  & $0.4$  & $0.102$ & $0.126$ & $3$ $(2)$   & $36$ $(30)$   & $6.22$ & $0.8041$ & $0.4010$ & $13\times10^{-4}$ \\
$0.6$  & $0.6$  & $0.126$ & $0.153$ & $2$ $(2)$   & $28$ $(23)$   & $7.33$ & $0.8575$ & $0.4080$ & $6\times10^{-4}$ \\
$0.8$  & $0.8$  & $0.162$ & $0.188$ & $1$ $(1)$   & $21$ $(17)$   & $8.93$ & $0.9039$ & $0.4035$ & $89\times10^{-4}$ \\
\hline
\end{tabular}
\end{center}
\label{tab:memory-vs-spin} 
\end{table*}

\section{Detectability of the memory modes}

The $m=0$ modes are small (on the order of $10\%$) compared to the
dominant $\ell=m$ components of the GW signal. However,
they exhibit both non-oscillatory growth and ringdown features, which
present some interesting aspects for detection and identification of
these modes. We discuss the prospects for observation in both
interferometers, and pulsar timing arrays, in the following sections.

\subsection{Interferometers}

The signal-to-noise ratio (SNR) within a given detector, $\rho$, in
GW searches based on matched filtering is given 
by~\cite{Flanagan:1998a}:
\begin{equation}
  \rho^2 \equiv \left(\frac{S}{N}\right)^2_{\rm matched}
    = 4 \int_0^\infty\frac{|\tilde{h}(f)|^2}{S_h(f)} df \,,
  \label{SignalToNoiseRatio}
\end{equation}
where $S_h(f)$ is the sensitivity curve of the instrument. Restricting
the integral to the $(\ell,m)=(2,0)$ mode, we can determine its
contribution to the overall SNR, testing against the proposed
advanced-LIGO~\citep{LIGO-sens-2010} and LISA noise
curves~\citep{LISAPETaskForce}.

Table~\ref{tab:memory-vs-spin} lists the maximum SNR attained
over a range of masses for a given model. For advanced-LIGO, the
optimal total mass is $M_\text{BH} \simeq 290M_\odot$
(independent of the model and invariant under the distance), i.e.~a
$145M_\odot+145M_\odot$ binary, for which we list the SNR at a
reference distance of $300\rm{Mpc}$ (the SNR scales linearly
with the distance).  The LISA results refer to an optimal (redshifted) total mass
of $M_\text{BH}=5.35\times 10^6M_\odot$ at a luminosity distance
$d=15.8{\rm Gpc}$ corresponding to a redshift $z=2$, where
models suggest at least one expected merger event per year
~\citep{Sesana:2004gf}.  By assuming a minimum SNR of
$\rho=8$ for detection, it will be possible to observe the
$(\ell,m)=(2,0)$ mode in LISA out to the given distance.
They are unlikely to be visible with significant event rates within
the planned advanced ground-based detectors, though future
generation experiments may have prospects.

\begin{figure}
\includegraphics[width=1.\linewidth]{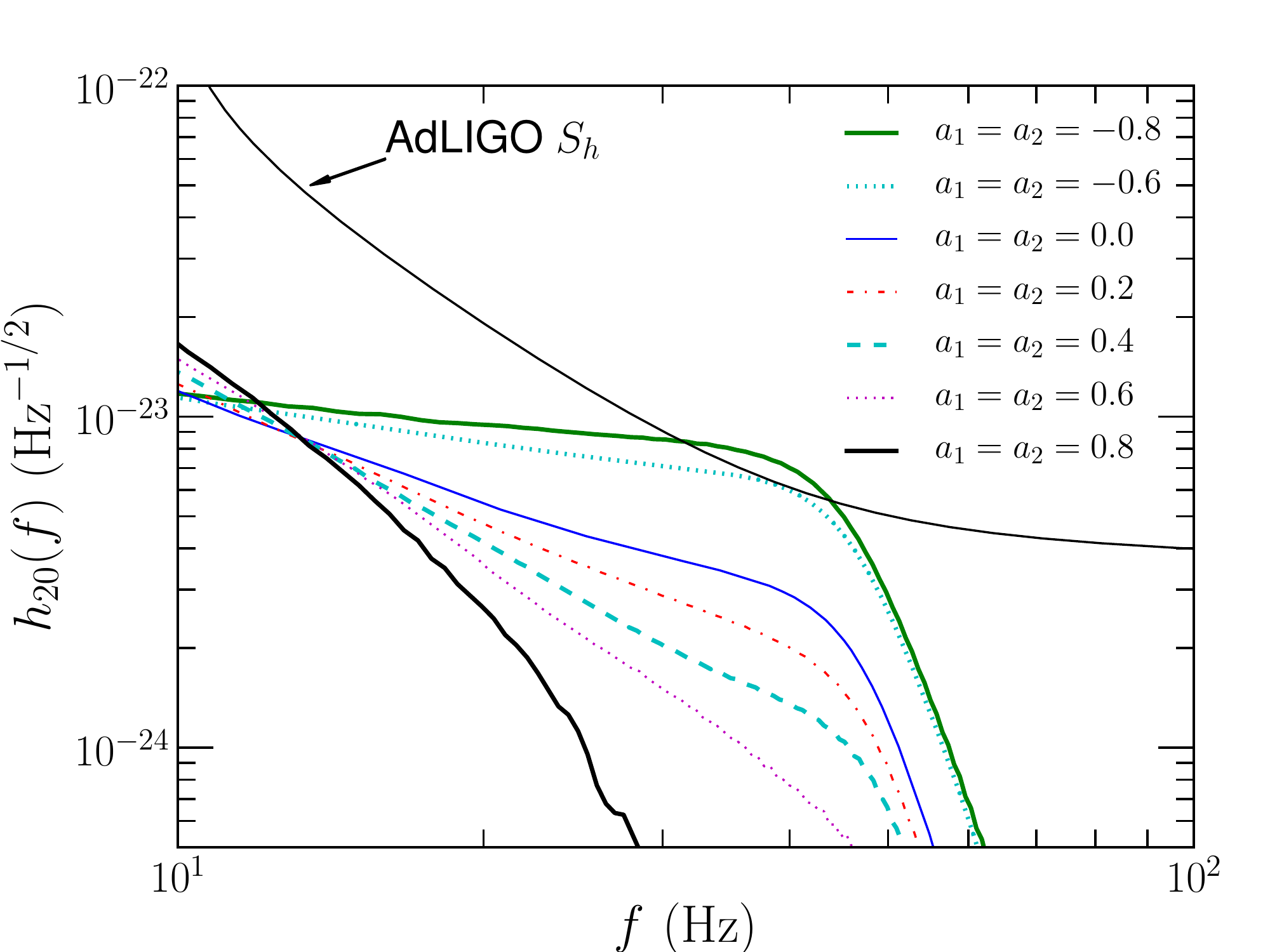}
  \caption{Fourier-representation of $(\ell,m)=(2,0)$ for different
    equal and aligned spin models at the optimal total mass $M=290M_\odot$
    ($145M_\odot+145M_\odot$ binary) at a distance $d=300{\rm Mpc}$
    and at an angle $\theta=\pi/2$. Models with higher total spin
    (e.g.~$a_1=a_2=+0.8$) have a weaker ringdown oscillation, which results in
    a higher overall slope and thus smaller amplitude at high
    frequencies as opposed to models with lower total spin
    (e.g.~$a_1=a_2=-0.8$).}
  \label{fig:snr-spin-dep}
\end{figure}

For both detectors, and as opposed to the dominant
$(\ell,m)=(2,2)$ mode \citep{Reisswig:2009vc}, we note that the highest SNR
occurs for the cases with spins \textit{anti-aligned} to the orbital
angular momentum which is due to their stronger ringdown.
In the \textit{aligned} cases, the weaker ringdown causes a 
much flatter high-frequency fall-off (Figure~\ref{fig:snr-spin-dep}), and this
effect overrides the non-oscillatory step within these detectors.

An important aspect of the matched-filtering algorithm is the
construction of templates against which signals are compared.  We
examine the influence of neglecting the $(2,0)$ contribution by
computing the mismatch between a signal, $h(\theta=\pi/3,\phi=\pi/2)$,
containing all of the $(2,m)$ modes, with one in which the $(2,0)$
mode is left out. For advanced-LIGO, the mismatch for the model
$a_1=a_2=+0.8$ is largest in the mass range $250M_\odot-400M_\odot$ ,
but negligible, with a magnitude of $10^{-11}$.  For the
$a_1=a_2=-0.8$ model, the mismatch grows to be on the order of
$10^{-5}$. Thus, although the $h_{20}$ mode induces a notable offset
in the wave signal when viewed in the time domain, it makes little
difference to the detection algorithms.

\subsection{Pulsar timing arrays}

An alternative proposal for measuring GWs may be
more sensitive to the non-oscillatory step-function nature of the
memory. Precise, distant clocks will experience a residual in
time-of-arrival measurements which can be associated with
GWs~\citep{1975GReGr...6..439E, Sazhin:1978gy,
  Detweiler:1979wn}.  Several such projects are in
development~\citep{Hobbs:2009yy, Manchester:2007mx, Jenet:2009hk,
  Lazio:2009bea}. While their principal aim is to examine the
stochastic background, recently a number of papers have pointed out
the potential for observing non-oscillatory step function burst
signals, such as the memory~\citep{Pshirkov:2009ak,
  vanHaasteren:2009fy, Seto:2009nv}.

Primary candidates for detection are supermassive BH binary
mergers. While there is a great deal of ongoing work understanding the
nature and evolution of such systems, if there is a tendency for the
bodies to align, then the results obtained here are directly
applicable. For instance, \cite{Dotti:2009vz} indicate that in
gas-rich mergers, the spins are aligned on a short timescale, with
spin magnitudes in the range $0.6$ to $0.8$.  Assuming a $m_1=m_2=10^8
M_\odot$ binary source at a distance of $1\text{Gpc}$ which has aligned spins
with a magnitude of $a_1=a_2=0.7$, the results of
Section~\ref{sec:spin} lead to an angle-averaged memory offset of:
\begin{equation}
  \langle h\rangle_\theta = 1.6 \times 10^{-16}
    \left( \frac{m}{10^8 M_\odot} \right)
    \left( \frac{1\text{Gpc}}{R} \right).
\end{equation}
(The corresponding constants for zero and maximally spinning bodies
are $0.9\times 10^{-16}$ and $2.0\times 10^{-16}$, respectively.)  The
results of~\cite{Pshirkov:2009ak} and~\cite{vanHaasteren:2009fy}
indicate that a burst will be observable over a 10-year observation
period at amplitude $h\simeq 2\times 10^{-15}$, suggesting that the
merger of a pair of $6\times 10^{8}$ with $a_1=a_2=0.7$ would be
observable at $1\text{Gpc}$.

\acknowledgements 
\textbf{Acknowledgments.} The authors would like to thank Marc Favata, Ian Hinder,
Sascha Husa and Christian D.~Ott for helpful input.  This work is
supported by the Bundesministerium f\"ur Bildung und Forschung and the
National Science Foundation under grant numbers AST-0855535 and
OCI-0905046. DP has been supported by grants CSD2007-00042 and
FPA-2007-60220 of the Spanish Ministry of Science.  Computations were
performed on the Teragrid (allocation TG-MCA02N014), the LONI network
(\texttt{www.loni.org}), at LRZ M\"unchen, the Barcelona
Supercomputing Center, and at the Albert-Einstein-Institut.



\begin{thebibliography}{50}
\expandafter\ifx\csname natexlab\endcsname\relax\def\natexlab#1{#1}\fi

\bibitem[{Alcubierre(2008)}]{Alcubierre:2008}
Alcubierre, M. 2008, Introduction to $3+1$ {N}umerical {R}elativity (Oxford,
  UK: Oxford University Press)

\bibitem[{Alcubierre {et~al.}(2003)Alcubierre, Br{\"u}gmann, Diener, Koppitz,
  Pollney, Seidel, \& Takahashi}]{Alcubierre02a}
Alcubierre, M., Br{\"u}gmann, B., Diener, P., Koppitz, M., Pollney, D., Seidel,
  E., \& Takahashi, R. 2003, Phys. Rev. D, 67, 084023

\bibitem[{Anninos {et~al.}(1995)Anninos, Hobill, Seidel, Smarr, \&
  Suen}]{Anninos94b}
Anninos, P., Hobill, D., Seidel, E., Smarr, L., \& Suen, W.-M. 1995, Phys. Rev.
  D, 52, 2044

\bibitem[{Babiuc {et~al.}(2005)Babiuc, Szil{\'a}gyi, Hawke, \&
  Zlochower}]{Babiuc:2005pg}
Babiuc, M., Szil{\'a}gyi, B., Hawke, I., \& Zlochower, Y. 2005, Class. Quantum
  Grav., 22, 5089

\bibitem[{Baker {et~al.}(2006)Baker, Centrella, Choi, Koppitz, \& van
  Meter}]{Baker:2005vv}
Baker, J.~G., Centrella, J., Choi, D.-I., Koppitz, M., \& van Meter, J. 2006,
  Phys. Rev. Lett., 96, 111102

\bibitem[{Berti {et~al.}(2009)Berti, Cardoso, \& Starinets}]{Berti:2009kk}
Berti, E., Cardoso, V., \& Starinets, A.~O. 2009, Class. Quant. Grav., 26,
  163001

\bibitem[{Bishop {et~al.}(1999)Bishop, Isaacson, G{\'o}mez, Lehner,
  Szil{\'a}gyi, \& Winicour}]{Bishop98b}
Bishop, N., Isaacson, R., G{\'o}mez, R., Lehner, L., Szil{\'a}gyi, B., \&
  Winicour, J. 1999, in Black {H}oles, {G}ravitational {R}adiation and the
  {U}niverse, ed. B.~Iyer \& B.~Bhawal (Kluwer, Dordrecht, The Neterlands), 393

\bibitem[{Blanchet {et~al.}(2006)Blanchet, Buonanno, \& Faye}]{Blanchet:2006gy}
Blanchet, L., Buonanno, A., \& Faye, G. 2006, Phys. Rev., D74, 104034

\bibitem[{Blanchet \& Damour(1992)}]{Blanchet:1992br}
Blanchet, L., \& Damour, T. 1992, Phys. Rev., D46, 4304

\bibitem[{Bondi {et~al.}(1962)Bondi, van~der Burg, \& Metzner}]{Bondi62}
Bondi, H., van~der Burg, M. G.~J., \& Metzner, A. W.~K. 1962, Proc. R. Soc.
  London, A269, 21

\bibitem[{{Braginskii} \& {Thorne}(1987)}]{1987Natur.327..123B}
{Braginskii}, V.~B., \& {Thorne}, K.~S. 1987, \nat, 327, 123

\bibitem[{Brandt \& Br{\"u}gmann(1997)}]{Brandt97b}
Brandt, S., \& Br{\"u}gmann, B. 1997, Phys. Rev. Lett., 78, 3606

\bibitem[{Campanelli {et~al.}(2006)Campanelli, Lousto, Marronetti, \&
  Zlochower}]{Campanelli:2005dd}
Campanelli, M., Lousto, C.~O., Marronetti, P., \& Zlochower, Y. 2006, Phys.
  Rev. Lett., 96, 111101

\bibitem[{Christodoulou(1991)}]{Christodoulou:1991cr}
Christodoulou, D. 1991, Phys. Rev. Lett., 67, 1486

\bibitem[{Detweiler(1979)}]{Detweiler:1979wn}
Detweiler, S. 1979, Astrophys. J., 234, 1100

\bibitem[{{Dotti} {et~al.}(2009){Dotti}, {Volonteri}, {Perego}, {Colpi},
  {Ruszkowski}, \& {Haardt}}]{Dotti:2009vz}
{Dotti}, M., {Volonteri}, M., {Perego}, A., {Colpi}, M., {Ruszkowski}, M., \&
  {Haardt}, F. 2009, Mon. Not. R. astr. Soc., 1795

\bibitem[{{Estabrook} \& {Wahlquist}(1975)}]{1975GReGr...6..439E}
{Estabrook}, F.~B., \& {Wahlquist}, H.~D. 1975, General Relativity and
  Gravitation, 6, 439

\bibitem[{Favata(2009{\natexlab{a}})}]{Favata:2008ti}
Favata, M. 2009{\natexlab{a}}, J. Phys. Conf. Ser., 154, 012043

\bibitem[{Favata(2009{\natexlab{b}})}]{Favata:2009ii}
---. 2009{\natexlab{b}}, Astrophys. J., 696, L159

\bibitem[{Favata(2009{\natexlab{c}})}]{Favata:2008yd}
---. 2009{\natexlab{c}}, Phys. Rev., D80, 024002

\bibitem[{Favata(2010)}]{Favata:2010zu}
---. 2010, arXiv:1003.3486

\bibitem[{Flanagan \& Hughes(1998)}]{Flanagan:1998a}
Flanagan, E.~E., \& Hughes, S. 1998, Phys. Rev. D, 57, 4535

\bibitem[{Goodale {et~al.}(2003)Goodale, Allen, Lanfermann, Mass{\'o}, Radke,
  Seidel, \& Shalf}]{Goodale02a}
Goodale, T., Allen, G., Lanfermann, G., Mass{\'o}, J., Radke, T., Seidel, E.,
  \& Shalf, J. 2003, in Vector and Parallel Processing -- VECPAR'2002, 5th
  International Conference, Lecture Notes in Computer Science (Berlin:
  Springer)

\bibitem[{Hannam {et~al.}(2007)Hannam, Husa, Pollney, Brugmann, \&
  O'Murchadha}]{Hannam:2006vv}
Hannam, M., Husa, S., Pollney, D., Brugmann, B., \& O'Murchadha, N. 2007, Phys.
  Rev. Lett., 99, 241102

\bibitem[{Hobbs {et~al.}(2009)}]{Hobbs:2009yy}
Hobbs, G., {et~al.} 2009, arXiv:0911.5206

\bibitem[{Jenet {et~al.}(2009)}]{Jenet:2009hk}
Jenet, F., {et~al.} 2009, arXiv:0909.1058

\bibitem[{Kennefick(1994)}]{Kennefick:1994nw}
Kennefick, D. 1994, Phys. Rev., D50, 3587

\bibitem[{Lazio(2009)}]{Lazio:2009bea}
Lazio, J. 2009, arXiv:0910.0632

\bibitem[{{LISA-wiki}(2010)}]{LISAPETaskForce}
{LISA-wiki}. 2010, {LISA} parameter estimation wiki, {\tt
  http://www.tapir.caltech.edu/dokuwiki/lisape:home}

\bibitem[{Lousto {et~al.}(2009)Lousto, Campanelli, \&
  Zlochower}]{Lousto:2009mf}
Lousto, C.~O., Campanelli, M., \& Zlochower, Y. 2009, arXiv:0904.3541

\bibitem[{Manchester(2008)}]{Manchester:2007mx}
Manchester, R.~N. 2008, AIP Conf. Proc., 983, 584

\bibitem[{Newman \& Penrose(1962)}]{Newman62a}
Newman, E.~T., \& Penrose, R. 1962, J. Math. Phys., 3, 566, erratum in J. Math.
  Phys. 4, 998 (1963)

\bibitem[{Payne(1983)}]{Payne:1983}
Payne, P.~N. 1983, Phys. Rev., D28, 1894

\bibitem[{Pollney {et~al.}(2009{\natexlab{a}})Pollney, Reisswig, Dorband,
  Schnetter, \& Diener}]{Pollney:2009ut}
Pollney, D., Reisswig, C., Dorband, N., Schnetter, E., \& Diener, P.
  2009{\natexlab{a}}, Phys. Rev., D80, 121502

\bibitem[{Pollney {et~al.}(2009{\natexlab{b}})Pollney, Reisswig, Schnetter,
  Dorband, \& Diener}]{Pollney:2009yz}
Pollney, D., Reisswig, C., Schnetter, E., Dorband, N., \& Diener, P.
  2009{\natexlab{b}}, arXiv:0910.3803

\bibitem[{Pshirkov {et~al.}(2009)Pshirkov, Baskaran, \&
  Postnov}]{Pshirkov:2009ak}
Pshirkov, M.~S., Baskaran, D., \& Postnov, K.~A. 2009, arXiv:0909.0742

\bibitem[{Reisswig {et~al.}(2009{\natexlab{a}})Reisswig, Bishop, Pollney, \&
  Szilagyi}]{Reisswig:2009us}
Reisswig, C., Bishop, N.~T., Pollney, D., \& Szilagyi, B. 2009{\natexlab{a}},
  Phys. Rev. Lett., 103, 221101

\bibitem[{Reisswig {et~al.}(2010)Reisswig, Bishop, Pollney, \&
  Szilagyi}]{Reisswig:2009rx}
---. 2010, Class. Quant. Grav., 27, 075014

\bibitem[{Reisswig {et~al.}(2009{\natexlab{b}})Reisswig, Husa, Rezzolla,
  Dorband, Pollney, \& Seiler}]{Reisswig:2009vc}
Reisswig, C., Husa, S., Rezzolla, L., Dorband, E.~N., Pollney, D., \& Seiler,
  J. 2009{\natexlab{b}}, \prd, 80, 124026

\bibitem[{Rezzolla {et~al.}(2008)}]{Rezzolla:2007xa}
Rezzolla, L., {et~al.} 2008, Astrophys. J, 679, 1422

\bibitem[{Sachs(1962)}]{Sachs62}
Sachs, R. 1962, Proc. Roy. Soc. London, A270, 103

\bibitem[{Sazhin(1978)}]{Sazhin:1978gy}
Sazhin, M.~V. 1978, Vestn. Mosk. Univ. Fiz. Astron., 19N1, 118

\bibitem[{Schnetter {et~al.}(2004)Schnetter, Hawley, \&
  Hawke}]{Schnetter-etal-03b}
Schnetter, E., Hawley, S.~H., \& Hawke, I. 2004, Class. Quantum Grav., 21, 1465

\bibitem[{Sesana {et~al.}(2005)Sesana, Haardt, Madau, \&
  Volonteri}]{Sesana:2004gf}
Sesana, A., Haardt, F., Madau, P., \& Volonteri, M. 2005, Astrophys. J., 623,
  23

\bibitem[{Seto(2009)}]{Seto:2009nv}
Seto, N. 2009, arXiv:0909.1379

\bibitem[{Shoemaker(2010)}]{LIGO-sens-2010}
Shoemaker, D. 2010, Advanced LIGO anticipated sensitivity curves, Tech. Rep.
  LIGO-T0900288-v3, LIGO Scientific Collaboration

\bibitem[{Vaishnav {et~al.}(2007)Vaishnav, Hinder, Herrmann, \&
  Shoemaker}]{Vaishnav:2007nm}
Vaishnav, B., Hinder, I., Herrmann, F., \& Shoemaker, D. 2007, Phys. Rev., D76,
  084020

\bibitem[{van Haasteren \& Levin(2009)}]{vanHaasteren:2009fy}
van Haasteren, R., \& Levin, Y. 2009, arXiv:0909.0954

\bibitem[{Winicour(2005)}]{Winicour05}
Winicour, J. 2005, Living Rev. Relativ., 8, 10, [Online article]

\bibitem[{{Zel'Dovich} \& {Polnarev}(1974)}]{1974SvA....18...17Z}
{Zel'Dovich}, Y.~B., \& {Polnarev}, A.~G. 1974, Soviet Astronomy, 18, 17

\end{thebibliography}
\end{document}